\documentclass[twocolumn,tighten]{aastex61}
\usepackage{xspace}

\newcommand{\source}{dm1647+21\xspace}
\newcommand{\HI}{H{\,{\sc i}}\xspace}
\newcommand{\Ha}{H{$\alpha$}\xspace}
\newcommand{\Msun}{M$_{\odot}$\xspace}
\newcommand{\pyr}{yr$^{-1}$\xspace}
\newcommand{\kms}{km~s$^{-1}$\xspace}
\newcommand{\cmc}{cm$^{-3}$\xspace}
\newcommand{\Hb}{H{$\beta$}\xspace}
\newcommand{\nii}{[N~\textsc{ii}]\xspace}
\newcommand{\oiii}{[O~\textsc{iii}]\xspace}
\newcommand{\oi}{[O~\textsc{i}]\xspace}
\newcommand{\sii}{[S~\textsc{ii}]\xspace}

\submitjournal{ApJ}

\shortauthors{Privon et al.}

\begin{document}

\correspondingauthor{G. C. Privon}
\email{gprivon@astro.puc.cl}

\title{A Widespread, Clumpy Starburst in the Isolated Ongoing Dwarf Galaxy merger dm1647+21}

\author[0000-0003-3474-1125]{G. C. Privon}
\affiliation{Instituto de Astrof\'sica, Facultad de F\'isica, Pontificia Universidad Cat\'olica de Chile, Casilla 306, Santiago 22, Chile}

\author{S. Stierwalt}
\affiliation{Department of Astronomy, University of Virginia, 530 McCormick Road, Charlottesville, VA 22904, USA}

\author{D. R. Patton}
\affiliation{Department of Physics and Astronomy, Trent University, 1600 West Bank Drive, Peterborough, ON K9L 0G2, Canada}

\author{G. Besla}
\affiliation{Department of Astronomy, University of Arizona, 933 North Cherry Avenue, Tucson, AZ 85721, USA}

\author{S. Pearson}
\affiliation{Department of Astronomy, Columbia University, 550 West 120th Street, New York, NY 10027, USA}

\author{M. Putman}
\affiliation{Department of Astronomy, Columbia University, 550 West 120th Street, New York, NY 10027, USA}

\author{K. E. Johnson}
\affiliation{Department of Astronomy, University of Virginia, 530 McCormick Road, Charlottesville, VA 22904, USA}

\author{N. Kallivayalil}
\affiliation{Department of Astronomy, University of Virginia, 530 McCormick Road, Charlottesville, VA 22904, USA}

\author{S. Liss}
\affiliation{Department of Astronomy, University of Virginia, 530 McCormick Road, Charlottesville, VA 22904, USA}

\collaboration{(TiNy Titans)}

\begin{abstract}
Interactions between pairs of isolated dwarf galaxies provide a critical window into low-mass hierarchical, gas-dominated galaxy assembly and the buildup of stellar mass in low-metallicity systems.
We present the first VLT/MUSE optical IFU observations of the interacting dwarf pair \source, selected from the TiNy Titans survey.
The \Ha emission is widespread and corresponds to a total unobscured star formation rate (SFR) of 0.44 \Msun \pyr, 2.7 times higher than the SFR inferred from SDSS data.
The implied specific SFR (sSFR) for the system is elevated by more than an order of magnitude above non-interacting dwarfs in the same mass range.
This increase is dominated by the lower-mass galaxy, which has a sSFR enhancement of $>$ 50.
Examining the spatially-resolved maps of classic optical line diagnostics, we find the ISM excitation can be fully explained by star formation.
The velocity field of the ionized gas is not consistent with simple rotation.
Dynamical simulations indicate that the irregular velocity field and the stellar structure is consistent with the identification of this system as an ongoing interaction between two dwarf galaxies.
The widespread, clumpy enhancements in star formation in this system point to important differences in the effect of mergers on dwarf galaxies, compared to massive galaxies: rather than the funneling of gas to the nucleus and giving rise to a nuclear starburst, starbursts in low-mass galaxy mergers may be triggered by large-scale ISM compression, and thus be more distributed.

\end{abstract}

\keywords{galaxies: ISM ---
galaxies: interactions ---
galaxies: starburst ---
galaxies: dwarf ---
galaxies: individual (SDSS J164710.66+210514.5,  SDSS J164711.12+210514.8)}

\section{Dwarf-dwarf Interactions and dm1647+21}
\label{sec:introduction}

Dynamical interactions and mergers between massive galaxies are known to effect dramatic structural changes in galaxies in addition to triggering starbursts and active galactic nuclei \citep[e.g.][]{Toomre1972,Barnes1988,Sanders1988,Stierwalt2013}.
These events are at least partially responsible for the morphological and color transformation of massive galaxies from late to early-type \citep[e.g.,][]{Hopkins2008}.
These interactions are well-studied, particularly at low-redshift.

However, comparatively little is known about the physical processes driven by interactions between low-mass galaxies.
Observations of low-redshift dwarf galaxy interactions and mergers probe hierarchical assembly in systems with low-metallicity gas and whose baryonic component is dominated by atomic gas.
The LMC/SMC pair provides a close-up view of an interacting pair of dwarfs, but the environment of the Milky Way makes it difficult to isolate the effects of the interaction \citep[e.g.,][]{Besla2012}.

In order to study galaxy assembly at low-mass, low-metallicity, and high gas fraction, we have identified a sample of SDSS-selected paired and unpaired dwarf galaxies, including both isolated systems and systems near a massive host.
The TiNy Titans \citep[TNT][]{Stierwalt2015} survey enables us to disentangle environmental effects (proximity to a massive host) and dynamical effects (interaction with a similar-mass companion) in the evolution of low-mass galaxies.
The paired sources are thought to be physically associated, based on their projected separations ($\Delta R<50$ kpc) and relative velocities ($\Delta v<300$ \kms).
The paired and unpaired sources are further separated into those isolated from a massive host (massive galaxies must be at $\Delta R>1.5$ Mpc, $\Delta v>1000$ \kms) and those associated with a massive host.

Fully understanding the influence mergers and interactions have on dwarf galaxies requires detailed studies of individual systems to spatially resolve the activity and dynamics.
We present the first results from Very Large Telescope (VLT) Multi-Unit Spectroscopic Explorer \citep[MUSE;][]{Bacon2010} observations of dwarf pairs.
Where appropriate, we adopt the WMAP9 cosmology \citep[H$_0=69.3$ \kms Mpc$^{-1}$, $\Omega_M=0.286$, $\Omega_{vacuum}=0.714$;][]{Hinshaw2013}.

\subsection{Overview of \source}

The object of the present study, \source, is a dwarf pair at 16h47m10.96s +21d05m14.5s with a redshift, $z=0.0090994$ (D$_L=39.6$ Mpc and $0.189$ kpc $/\arcsec$).
The pair consists of the SDSS spectroscopic objects SDSS J164710.66+210514.5 and SDSS J164711.12+210514.8 which have a projected linear separation of $5$ kpc and a velocity difference of $27$ \kms \citep{Stierwalt2015}.
The system also meets the above isolation criteria from a massive galaxy.
Using the multi-band SDSS photometry, we compute a total stellar mass for the system of $9.0\times10^{7}$ \Msun \citep[calculated as described in][]{Stierwalt2015} with a mass ratio between the two components of 4.6:1.
Based on the SDSS spectroscopic measurements and an extrapolation to the entire galaxy the system has an estimated star formation rate of 0.15 \Msun \pyr \citep{Brinchmann2004}.
The metallicity inferred for the system is Z$=7.84$ \citep{Tremonti2004}.

The pair has a nearby companion galaxy at a similar redshift, UGC~10549, which has a stellar mass of $2.4\times10^{8}$ \Msun \citep[from SDSS DR12, using the ][ method]{Maraston2006}.
UGC~10549 does not formally meet the association criterion for TNT pairs, with a projected separation to \source just over $50$ kpc and thus was not initially identified as being associated.
However, considering the proximity of this object to \source, this may be a potential dwarf-only group, similar to those found by \citet{Stierwalt2017}.

Using integral field unit (IFU) observations of \source (Section~\ref{sec:observations}) we explore the activity and kinematics of the system (Section~\ref{sec:ism}) to test the dwarf-dwarf interaction scenario for the system and compare the activity to those of massive galaxy mergers (Section~\ref{sec:ddmerger}).

\section{MUSE IFU Observations of \source}
\label{sec:observations}

A total of 2895 seconds of on-source integration were obtained for \source with MUSE in clear conditions on 2016 Sept.\ 05 in three exposures with the Wide Field Mode and nominal wavelength coverage ($480-930$ nm).
The data were reduced and calibrated with the ESO Reflex pipeline software \citep{Freudling2013}.
A sky background was subtracted using emission-free regions of the cube.

The 1 arcminute extent of the MUSE FOV corresponds to a linear scale of 11.3 kpc at the distance of \source, and covers the full optical extent of the system.
The cube has a resolution of $\sim0.90\arcsec$ (170 pc) and $\sim0.75\arcsec$ (140 pc) at 500 nm and 800 nm, respectively.
The spectral resolution varies from $R=1750-3750$, from the blue to the red.

\subsection{Creation of Broadband and Line Images}

\begin{figure*}
\centering
\includegraphics[width=\textwidth]{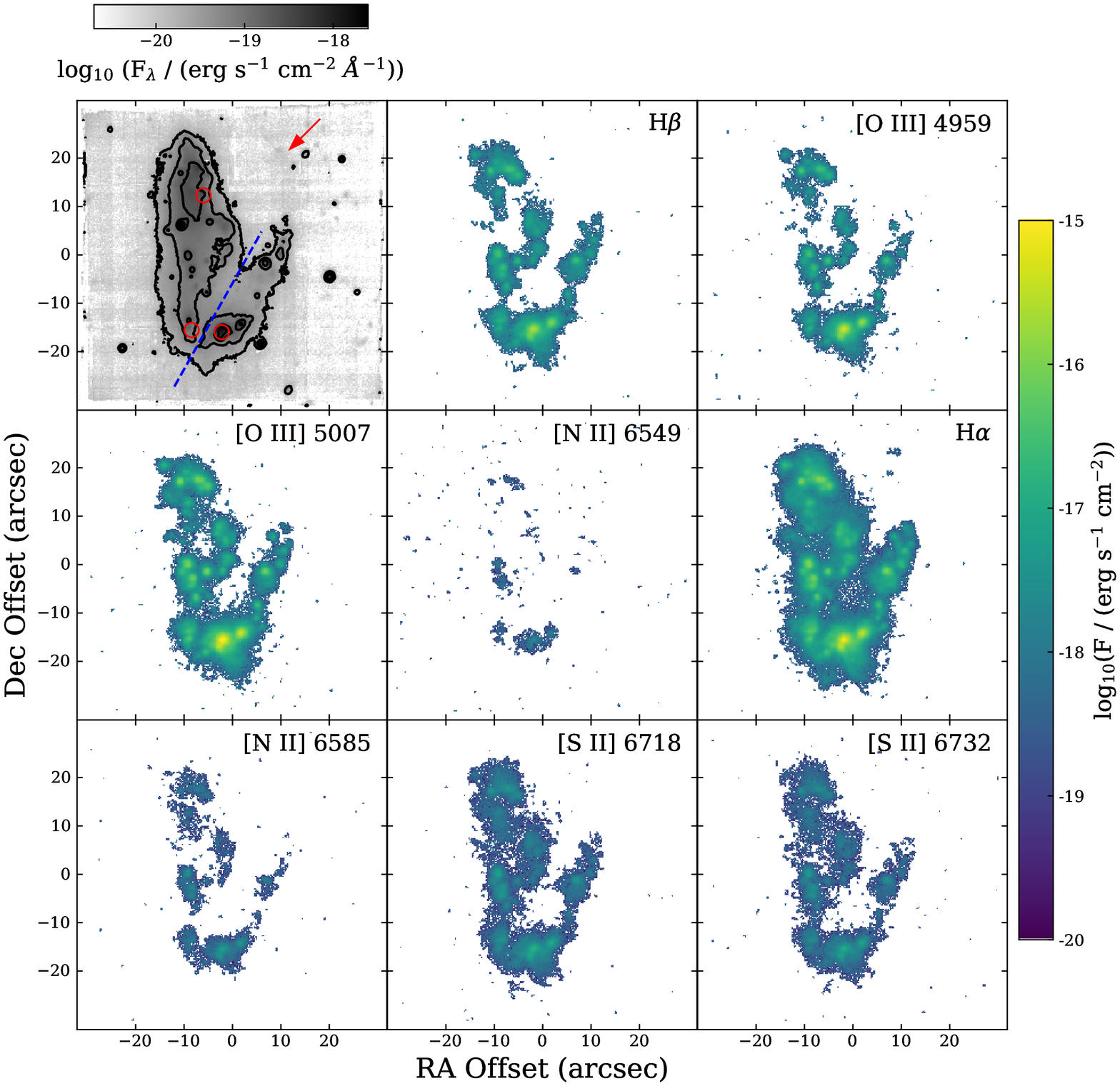}
\caption{Images created from the MUSE data cube.
Top row, from left: \emph{r} broadband image, \Hb, and \oiii 4959.
Middle row, from left: \oiii 5007, \nii 6549, and \Ha.
Bottom row, from left: \nii 6585, \sii 6718, and \sii 6732.
All images span the same field of view (1 arcmin, 11.3 kpc on a side) centered on 16h47m10.59s $+$21d05m32.15s.
The red circles in the upper-left panel mark the approximate location and size of available SDSS fibers; the Northern and Southwestern fibers were used in the identification of the pair.
The arrow in the upper-left panel marks the location of low-significance emission.
This is suggestive of faint tidal debris extending from the southern galaxy and curving up towards the north, which is consistent with a tidal tail from the lower-mass galaxy.
The \Ha image shows two clumps of emission at a similar location, further consistent with the possible identification of tidal material.
The blue dashed line over the \emph{r} band image marks the approximate dividing line adopted to separate the two galaxies comprising the \source system.
The line images are all shown on the same logarithmic intensity scale and are not corrected for extinction.
}
\label{fig:lines}
\end{figure*}

Using the MUSE datacube and the SDSS \emph{r} filter transmission curve, we construct a broadband image for \source (Figure~\ref{fig:lines}, upper left).
We also detect emission from the important optical diagnostic lines: \Hb, the \oiii doublet, \Ha, the \nii doublet, and the \sii doublet (Table~\ref{table:fluxes}, Figure~\ref{fig:lines}).
The \oi line at 6300\AA\ was redshifted to the location of a sky line and we cannot identify a clear detection.
Line images were created by using adjacent line-free channels to subtract the continuum underneath the line, then summing the channels visually identified to contain line emission.
The resulting images were masked by convolving each image with a Gaussian with $\sigma=1.5$ pixels and determining the RMS value of an emission-free portion of the image.
Pixels in the unsmoothed line image with values less than three times this RMS value were masked.

\begin{deluxetable}{lDD}
\tablecaption{Line Detections}
\tablehead{
\colhead{Line} & \multicolumn2c{Integrated Flux} & \multicolumn2c{Uncertainty} \\
& \multicolumn4c{($\times10^{-15}$ erg s$^{-1}$ cm$^{-2}$)}}
\decimals
\startdata
\Hb & 55.02 & 0.07 \\
\oiii 4959 & 59.70 & 0.07 \\
\oiii 5007 & 178.47 & 0.07  \\
\nii 5755 & <0.15 & . \\
\nii 6549 & 0.94 & 0.04 \\
\Ha & 189.22 & 0.05 \\
\nii 6585 & 3.25 & 0.03 \\
\sii 6718 & 11.30 & 0.03 \\
\sii 6732 & 8.13 & 0.03 \\
\enddata
\tablecomments{Line fluxes were measured from total intensity images (Figure~\ref{fig:lines}) and errors were computed using the per-pixel variance values calculated during the pipeline reduction.
The upper limit value quoted for \nii 5755 is $3\sigma$.}
\label{table:fluxes}
\end{deluxetable}

\section{Optical Properties of \source}
\label{sec:ism}

\subsection{Optical Continuum Emission}

The deep \emph{r}-band image generated from the MUSE data shows low-significance extended emission which may be an extension of tidal material from the lower-mass galaxy.
The background level of the image is dominated by striations which likely result from imperfect illumination correction for the IFU modules.
These limit our per-pixel $3\sigma$ background sensitivity to $\mu_r\sim25.5$ mag arcsec$^{-2}$ ($\sim2$ mag arcsec$^{-2}$ fainter than the SDSS \emph{r} image for this field).
At this level, we see no evidence for an underlying envelope of extended continuum emission that would suggest this is a single, irregular galaxy.
Future deep optical continuum observations \citep[e.g.,][]{Duc2015} would be useful in confirming the possible faint tidal emission and searching for evidence of an extended stellar envelope.
The blue dashed line in Figure~\ref{fig:lines} shows the approximate dividing line adopted for the photometric separation of the two component galaxies.

\subsection{Star Formation}
\label{sec:SF}

From the system-integrated \Ha flux, we compute a \Ha luminosity of $3.5\times10^{40}$ erg s$^{-1}$.
Assuming an intrinsic \Ha/\Hb ratio of 2.86 and utilizing the LMC average extinction curve from \citet{Gordon2003}, we compute A$_{H\alpha}\sim0.2-1.4$, with a system-averaged value of 0.51.
\footnote{The extinction and the subsequent extinction-corrected SFRs are relatively robust to the choice of a LMC versus MW extinction curve. Using a MW extinction curve as in \citet{Lee2009} produces values within $10$\% of those presented here.}
Using this calculated extinction value, the intrinsic \Ha luminosity is $5.6\times10^{40}$ erg s$^{-1}$.
Using the \citet{Kennicutt1998} relation for the star formation rate, we compute a SFR of 0.44 \Msun \pyr (corrected for extinction) for the system, a factor of 2.7 higher than the total SFR estimated from the extrapolation of the SDSS spectroscopy to the whole system \citep[e.g.,][]{Brinchmann2004}.
The observed equivalent width of the \Ha emission is $118$ \AA, so this system belongs to the population of starbursting dwarfs \citep[\Ha EQW $> 100$ \AA;][]{Lee2009a}.

The \source pair was detected in WISE, with a W4 ($22~\mu m$) flux of 6.43 mJy \citep{Wright2010}.
If this flux is entirely due to star formation, this corresponds to an obscured SFR of $0.04$ \Msun \pyr \citep[using the calibration from][]{Chang2015a}, an order of magnitude smaller than the SFR inferred from \Ha.
This indicates that the majority of the star formation is occurring in regions not heavily obscured by dust and is consistent with the relatively low extinction inferred from the Balmer decrement.
Thus, we are confident that the \Ha image is indicative of the true distribution of star formation in \source.
Summing the obscured and unobscured SFR estimates, we find a total SFR of $0.48$ \Msun \pyr.
Considering the measured stellar mass from SDSS photometry and the total SFR, the specific SFR (sSFR) is $5.3\times10^{-9}$ \pyr, corresponding to a mass-doubling time of 190 Myr.
The northern galaxy has $\sim40\%$ of the \Ha flux and the southern system has the remainder.
Considering the 4.6:1 stellar mass ratio, the sSFR for the lower mass galaxy is a factor of 7 larger than the more massive galaxy.

The brightest \Ha clump in the system has an observed \Ha flux of $23.3\times10^{-15}$ erg s$^{-1}$ cm$^{-2}$ (measured within a $1.5\arcsec$ circular aperture), accounting for 12\% of the total \Ha flux for the system.
The two next most luminous clumps have extinction-corrected fluxes of $10.6$ and $9.9\times10^{-15}$ erg s$^{-1}$ cm$^{-2}$.
Together these three clumps---all found in the lower-mass galaxy---account for $\sim23\%$ of the \Ha luminosity.
The remaining $\sim30$ clumps are distributed across both galaxies and each individually contribute less than $\sim2\%$ to the total flux; these clumps, and the remaining diffuse emission, effectively fill the same regions where we see stellar continuum emission.
This clump analysis is based on visual identification of pointlike emission in the \Ha image; a more detailed analysis of the clumps in \source and other systems will be presented by S. Liss et~al.\ (2017, \emph{in preparation}).

\subsubsection{On the Gas Depletion Time}
\label{sec:depletion}

Single-dish \HI observations of the system show a gas mass of $3.7\times10^{9}$ \Msun \citep{Stierwalt2015}, however this measurement likely includes \HI associated with a nearby system that is within the 8 arcminute GBT beam but not part of the interacting pair (UGC~10549, $z=0.009$, 50 kpc projected separation).
The integrated spectrum shows a somewhat complex profile, with the velocity of the peak emission at $\gtrsim50$ \kms lower velocity than the redshifts of the two components of \source.
This peak corresponds to the redshift of UGC~10549, suggesting a substantial amount of the single-dish \HI flux is associated with that galaxy.

In support of this interpretation of the single-dish profile, follow-up interferometric observations with the VLA reveal a clumpy \HI distribution in \source and emission associated with UGC~10549 (S. Stierwalt et~al.\ 2017 \emph{in preparation}).
Unfortunately, the VLA observations (taken in the B-array) do not recover the total single-dish \HI flux, so we cannot reliably separate the single-dish \HI mass into contributions from \source and UGC~10549.
Considering the uncertain distribution of the \HI mass, the single-dish measurement implies an \emph{upper limit} to the gas depletion time of $2$ Gyr, considering only the SFR of \source.
The true gas depletion time for \source is likely shorter than this estimate.

\subsection{ISM Properties}
\label{sec:ISMproperties}

\begin{figure*}
\centering
\includegraphics[width=\textwidth]{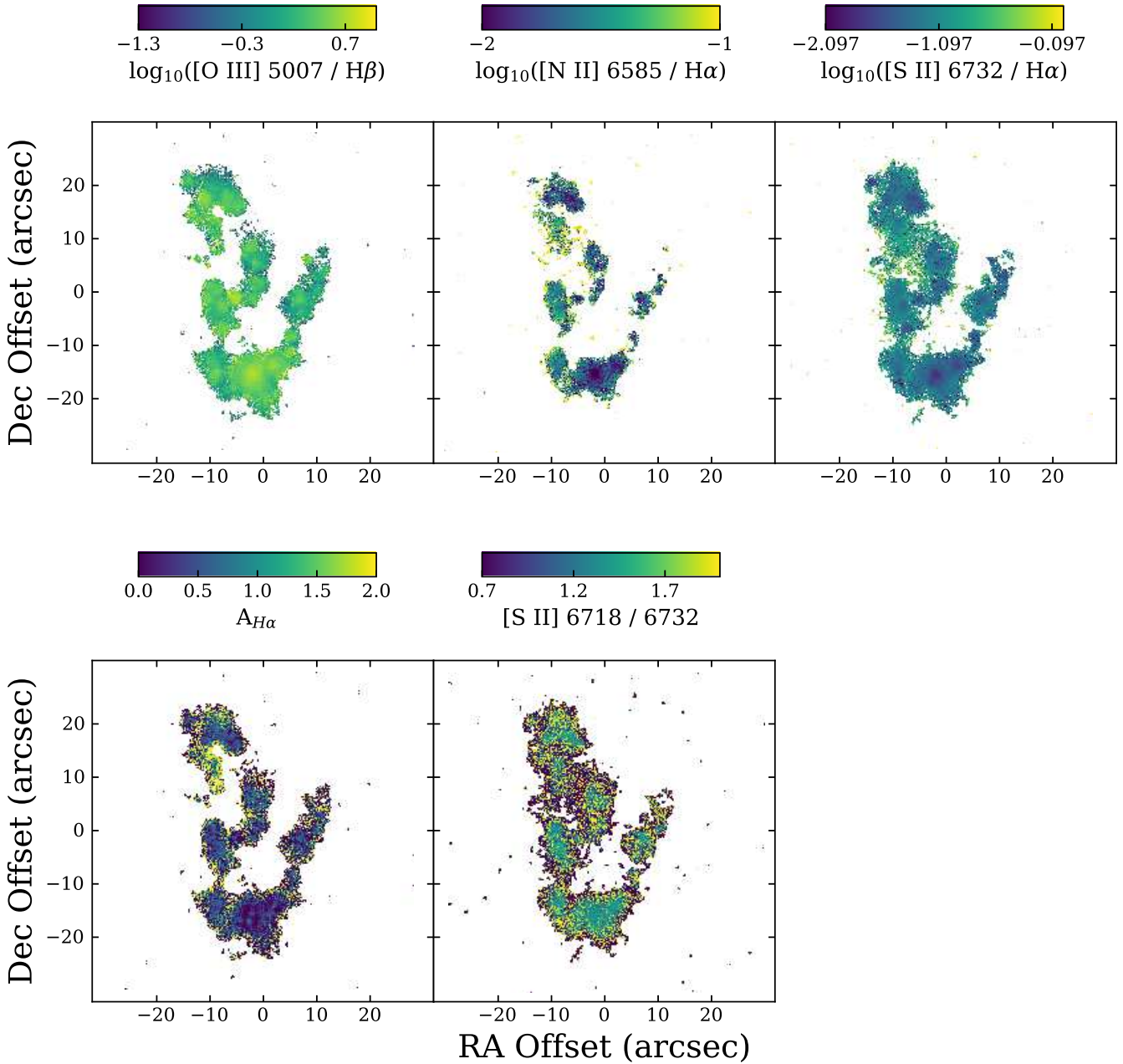}
\caption{Optical line ratio images. First row, from left to right: \oiii / \Hb, \nii / \Ha, \sii / \Ha.
Second row, from left to right: A$_{H\alpha}$ and \sii 6718 / \sii 6732.
The ratio maps were all masked to the fainter of the two lines (Figure~\ref{fig:lines}).
No extinction corrections have been applied.
}
\label{fig:ratios}
\end{figure*}

\begin{figure*}
\centering
\includegraphics[width=\textwidth]{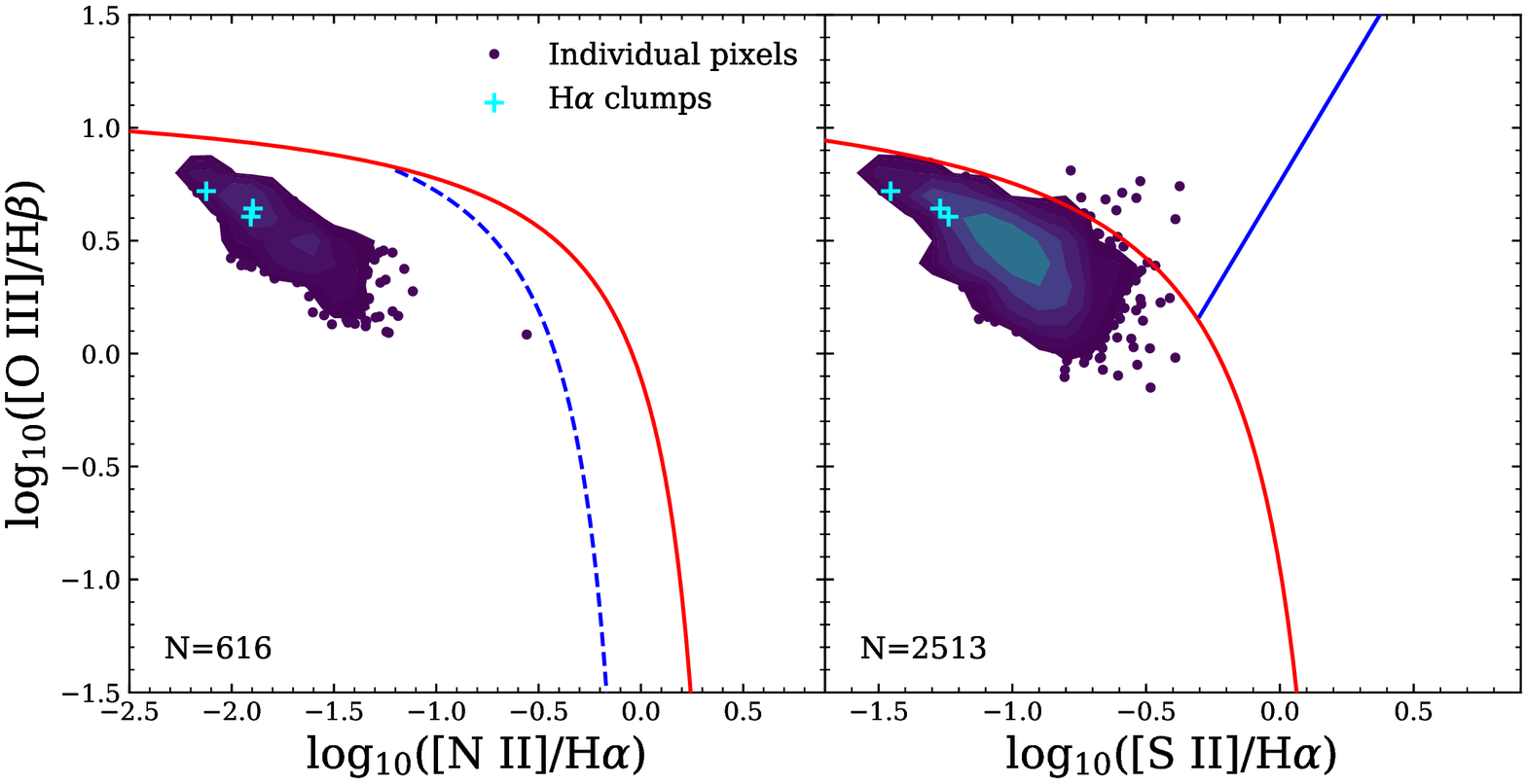}
\caption{
Optical line diagnostic diagrams based on the narrow-band images in Figure~\ref{fig:ratios}.
Left: \oiii/\Hb versus \nii/\Ha.
Right: \oiii/\Hb versus \sii/\Ha.
The colored contours and small points show the distribution of values for individual $0.2\arcsec\times0.2\arcsec$ spaxels in the datacube; the number in the lower left gives the number of spaxels shown in each panel.
The plotted spaxels are those with a $\geq5\sigma$ detection in \nii 6585 (left) or \sii 6718 + 6732 (right).
The cyan crosses mark the locations of the three brightest clumps identified from the \Ha map (Section~\ref{sec:SF}, $1.5\arcsec$ diameter apertures).
The colored lines are the \citet{Kewley2006} separations between star forming galaxies, Seyferts, and LINERs.
The line ratios are consistent with excitation from ongoing star formation.
}
\label{fig:BPT}
\end{figure*}

We use the line images created from the MUSE data to construct optical line ratio maps, based on classic optical diagnostic plots \citep[e.g.,][]{Baldwin1981,Veilleux1987,Kewley2006}.
Maps of \oiii/\Hb, \nii/\Ha, and \sii/\Ha are shown in Figure~\ref{fig:ratios}.
The system-integrated values for these ratios are: $\log_{10}($\oiii/\Hb$)=0.51$, $\log_{10}($\nii/\Ha$)=-1.8$, and $\log_{10}($\sii/\Ha)$=-1.0$.
In Figure~\ref{fig:BPT} we show the line diagnostic diagrams with the distribution of values from Figure~\ref{fig:ratios} and the source separation lines from \citet{Kewley2006}.
The line ratios in the ratio maps and diagnostic diagrams are all consistent with excitation from star formation and we see no evidence for LINER-like or Seyfert emission.

Using the system-integrated \sii line ratio of $1.40$, we estimate an upper limit to the electron density, $n_e<10^{2}$ \cmc \citep{Osterbrock2006}.
The spatial variation in this ratio is minimal in the regions with high signal-to-noise (Figure~\ref{fig:ratios}), suggesting consistently low $n_e$ across the source.
This low value may be due to a high-filling factor of low-density ionized gas and a low filling factor of young star clusters (that would have higher $n_e$ densities).

We measure a system-integrated \nii 6585 / \nii 6549 ratio of $3.5$, somewhat higher than the ratio fixed by quantum mechanics.
This is likely due to the 6585 flux including regions where 6549 fell below the detection threshold, artificially inflating the global ratio.
The \nii 5755 line is not detected, and we estimate a $3\sigma$ upper limit of  $0.15\times10^{-15}$ erg s$^{-1}$ cm$^{-2}$.
Combining this with the sum of the two detected lines, the \nii emission is consistent with a temperature of $T_{nebula}<20,000$~K.

\subsection{ISM Kinematics}

\begin{figure*}
\centering
\includegraphics[width=\textwidth]{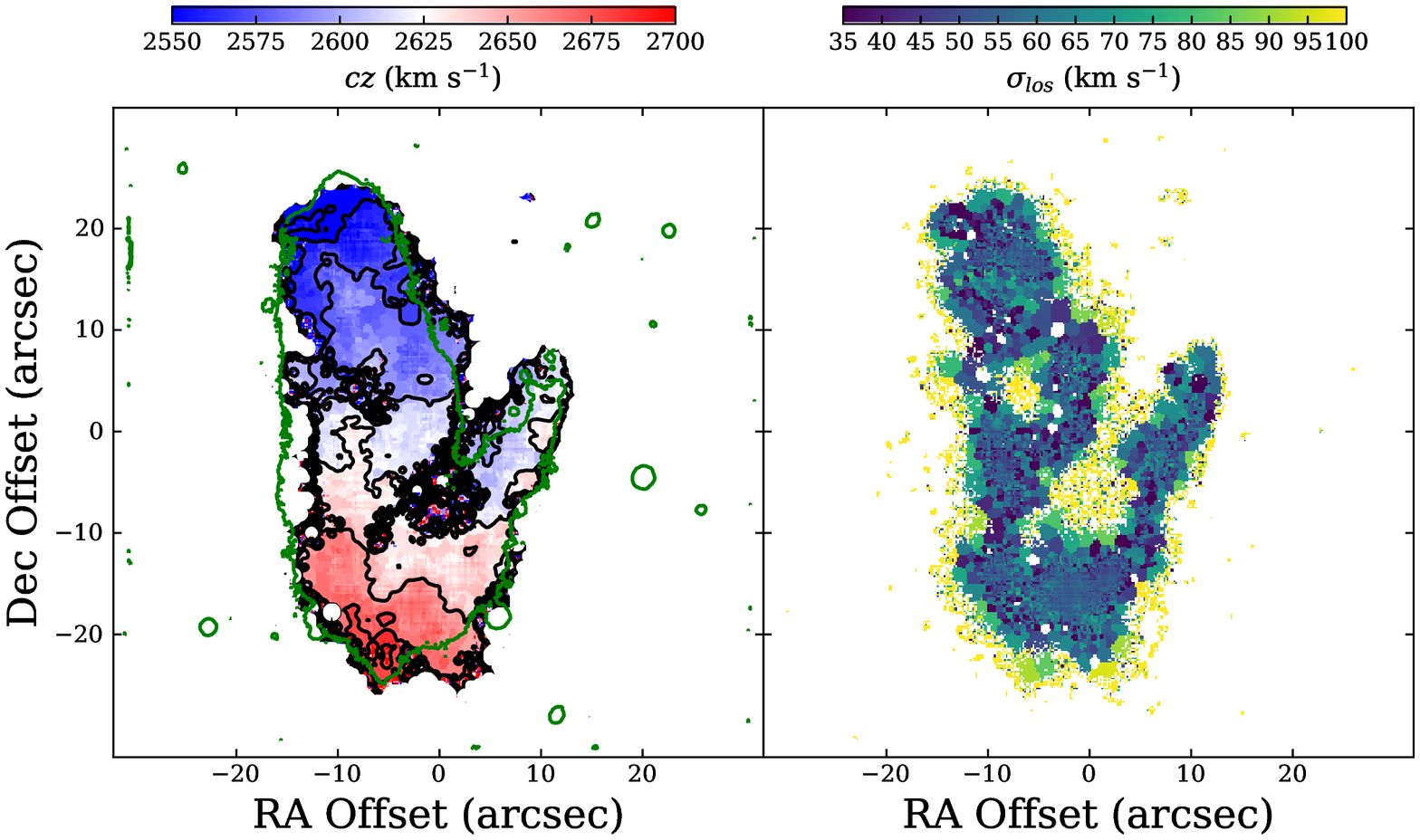}
\caption{Left: Intensity-weighted line of sight velocity from the \Ha emission shown as a function of position.
The black contours are spaced at $25$ \kms intervals between 2550--2700 \kms.
The green contour is the lowest contour from the $r$ image in Figure~\ref{fig:lines}.
Right: Intensity-weighted velocity dispersion of \Ha, shown as a function of position.
The velocity dispersion map is consistent with \Ha being spectrally unresolved.
}
\label{fig:kinematics}
\end{figure*}

In Figure~\ref{fig:kinematics} we show the intensity-weighted line-of-sight velocity and dispersion for the \Ha line.
The images were constructed by Voronoi binning \citep[e.g.,][]{Cappellari2003} the \Ha line image (Figure~\ref{fig:lines}) to a signal-to-noise of 50, including only individual pixels with $\geq3\sigma$.
The resulting binning scheme was then applied to a continuum-subtracted data cube containing \Ha and the intensity-weighted moment maps were computed from this binned cube.
This velocity field is consistent with the pair separation of 27 \kms, considering the locations of the SDSS fibers.
The system shows a clear velocity gradient from north-to-south, spanning a total of $\sim150$ \kms.
The \Ha velocity field for \source shows irregular iso-velocity contours, rather than the classic ``spider'' pattern of a smoothly rotating disk \citep[e.g.,][for dwarf galaxies]{Swaters2002}.
Due to the isolation of \source from a nearby massive galaxy, external perturbation on a single object are unlikely to be the origin of this complex kinematic structure.
The kinematics suggests the ionized gas is not part of a large, regularly rotating disk but instead are consistent with the system being a dynamical interaction of two dwarf galaxies.

The velocity dispersion is consistent with the instrumental resolution ($\lesssim100$ \kms), across the entire source, and we see no evidence for broadened lines indicative of shocks.
This, as well as the line ratio maps (Figure~\ref{fig:ratios}) suggests large-scale shocks from the ongoing merger are not contributing substantially to the energetics of the ISM.
If the line-of-sight velocities are representative (i.e., the bulk of the motion of these galaxies is not in the plane of the sky), they suggest a relatively slow encounter, compared to major interactions between massive galaxies.

\section{\source as a Dwarf-Dwarf Merger}
\label{sec:ddmerger}

Based on the velocity field of \Ha and optical morphology , \source is consistent with being a pair of interacting dwarf galaxies.
In the following two sub-sections we briefly explore a ``lookalike'' dynamical model for the system and contrast the observed activity in \source with the general characteristics of massive galaxy interactions.

\subsection{A ``Lookalike'' Numerical Model}

In the spirit of \citet{Toomre1972}, we created a ``lookalike'' dynamical model for this system, using the Identikit dynamical modeling tool \citep{Barnes2009}.
While a bona fide dynamical model would need to be constrained by kinematics of tidal features or be verified by a simulation employing a self-gravitating disk, this lookalike model is instructive for qualitatively demonstrating the merger scenario for \source.
We explored Identikit models corresponding to 2:1 mass ratio encounter on a parabolic orbit (e=1).
The galaxy models are the same as those used in \citet{Barnes2009} and \citet{Privon2013}, with the scale lengths of the half-mass galaxy reduced by a factor $\sqrt{2}$ to maintain a constant mass surface density.
While the observed stellar mass ratio of the two galaxies is $4.6:1$, the baryonic mass ratio of the progenitors is uncertain owing to the likely dominant contribution of \HI over the stellar mass.
The lower-mass galaxy is likely to have a higher gas fraction \citep[e.g.,][]{Swaters2002,Bradford2015}, implying a baryonic mass ratio that could be closer to 2:1.

\begin{figure}
\centering
\includegraphics[width=0.45\textwidth]{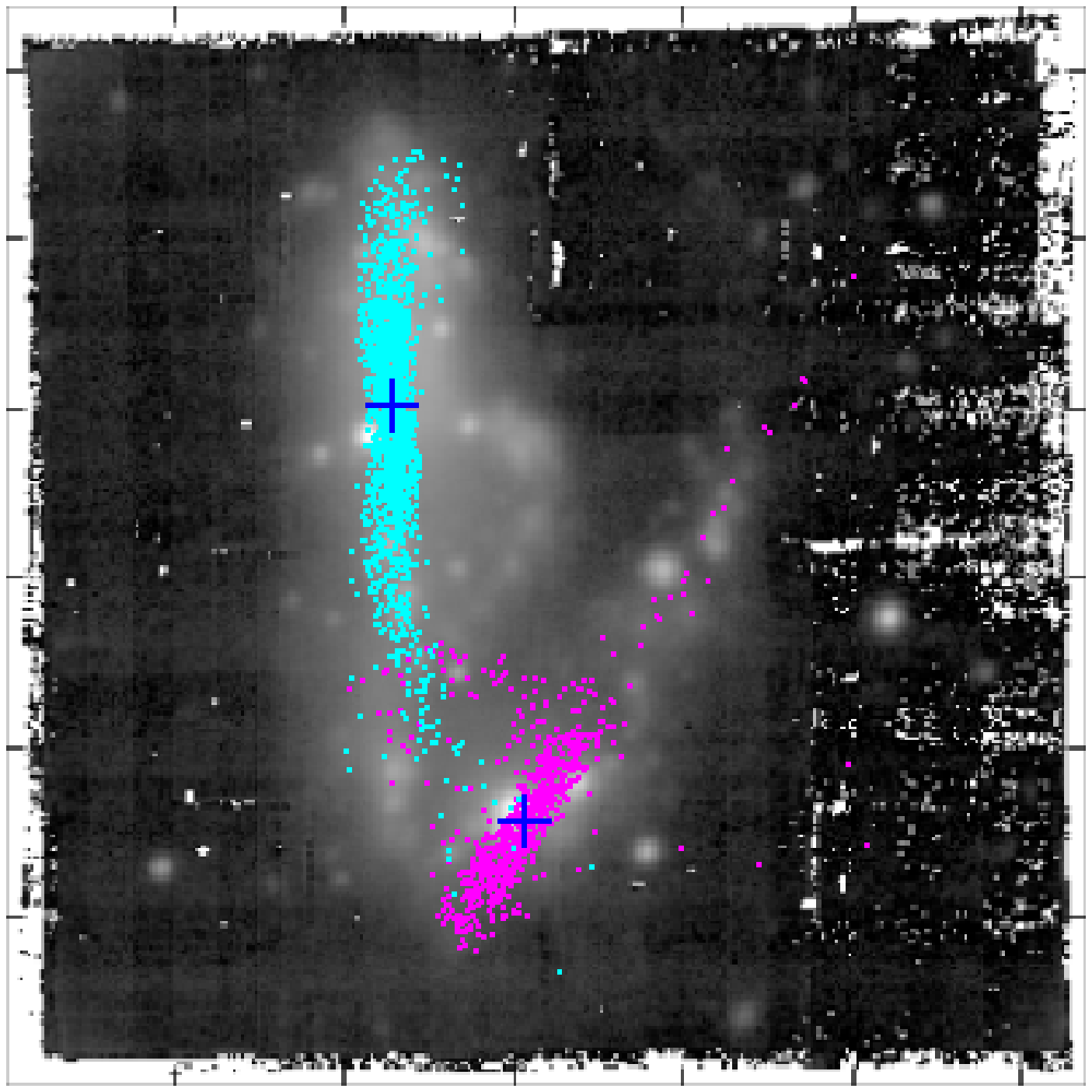}
\caption{\emph{r} band image, overlaid with a lookalike dynamical model.
The cyan and magenta points mark the positions of massless test particles from the disks of the two progenitor galaxies.
The general morphology of the system can be explained by a 2:1 mass ratio with a retrograde (more massive galaxy; cyan) -- prograde (less massive galaxy; magenta) encounter being viewed shortly after the first close pass of the disks.
}
\label{fig:model}
\end{figure}

We manually explored a range of initial galaxy orientations (inclination of the disks relative to the orbital plane), time since first close pass, and viewing directions relative to the orbital plane
\footnote{For a detailed description of the Identikit modeling process, see \citet{Barnes2009} and \citet{Privon2013}.}.
We were able to reproduce the overall morphology and general sense of the velocity gradient using a 2:1 mass ratio encounter where the more massive galaxy has a retrograde spin, the less massive galaxy has a prograde spin, and the system is being viewed after the first close pass of the two disks.
In Figure~\ref{fig:model} we show an overlay of the massless test particles in our lookalike model with the grayscale \emph{r} band image.
The major features of the system can be qualitatively explained by this ``lookalike'' modeling, lending additional support to the identification of \source as a pair of interacting dwarf galaxies.
We additionally explored encounters with a total mass ratio of 4:1 (motivated by the observed stellar mass ratio) and were able to reproduce the observed morphology with a model in which both galaxies experience a prograde interaction.
The morphology of the 4:1 lookalike is broadly consistent with that of the 2:1 mass ratio encounter, owing to the reduced tidal effect of the lower-mass galaxy on the massive companion.
Distinguishing between these encounter scenarios will require detailed mapping of the total atomic gas distribution and subsequent detailed modeling of tidal features.
Based on the small projected velocity range across the entire source, it seems likely these two dwarf galaxies are bound and will ultimately merge into a single dwarf.

In future TNT papers we will explore detailed dynamical modeling of individual observed dwarf-dwarf interactions (including constraints from spatially-resolved kinematics; S. Pearson et~al.\ 2017 \emph{in preparation}) as well as the properties of simulated paired and unpaired dwarfs in large-volume cosmological simulations (G. Besla et~al.\ 2017 \emph{in preparation}).

\subsection{Comparison with Massive Galaxy Mergers}

\citet{Stierwalt2015} showed that isolated dwarf pairs show star formation rates enhanced over isolated single dwarfs, with that enhancement increasing with decreasing pair separation.
Massive galaxies show similar enhancements \citep{Patton2013}, however the SFR enhancements in massive mergers tend to occur in the nuclei \citep[e.g.,][]{Armus1989,Mihos1994a,Barnes1996} and SF may even be suppressed on large scales due to the inflow of gas to the nucleus  \citep{Moreno2015}.

In contrast, the star formation in \source is both enhanced and distributed across the source, not concentrated in one or two regions (i.e., there are many, distributed clumps and no single clump dominates the total SFR).
The sSFR for this system is approximately a dex above that typical for dwarf galaxies in this mass range \citep[sSFR $\sim 3\times 10^{-10}$;][]{Lee2011}, consistent with the starburst identification of this system, based on the \Ha EQW.
Considering the lower-mass galaxy alone, its sSFR is a factor of $>50$ in excess of typical galaxies of that mass.
So while the distributed, clumpy star formation is morphologically similar to other dwarf galaxies of similar mass, the significant enhancement points to the ongoing interaction as the likely driver of enhanced star formation in \emph{both} galaxies.
By comparison, the sSFR of \source exceeds that of low-redshift infrared-selected luminous and ultraluminous infrared galaxies \citep{Howell2010}.

Why is the strongly enhanced star formation less concentrated in \source, compared to more massive galaxy mergers?
A clue may lie in the typical \HI properties of dwarfs.
The baryonic mass of these dwarf pairs is dominated by the \HI \citep{Stierwalt2015} and in a \HI study of a nearby sample of dwarf pairs, \citet{Pearson2016} found a significant fraction ($20-60\%$) of the neutral atomic gas lies beyond the optical extent of these dwarf pairs.
Extended \HI distributions are also observed for non-interacting dwarf galaxies \citep{Swaters2002}.
It thus seems generic for dwarf galaxies to have large-scale \HI envelopes.

\citet{Jog1992} proposed a scenario for the triggering of star formation in galaxy merger overlap regions in which the collision of \HI clouds increases the pressure in the cloud.
The increased pressure within the \HI envelope in turn triggers the collapse of pre-existing, formerly stable molecular clouds embedded within the \HI, resulting in a starburst.
The elevated yet distributed starburst in \source may be a consequence of the interaction of pre-existing \HI envelopes of the two progenitor galaxies triggering the collapse of molecular clouds across the system.
This is in contrast to massive galaxy mergers, where gas funneling to the nucleus is generally thought to precede the onset of the starburst \citep[e.g.,][]{Barnes1991,Barnes1996}.

Thus, if \source is a typical dwarf-dwarf merger the interactions enhance star formation but perhaps due to over-pressuring of \HI envelopes instead of funneling gas to the nuclei.
A consequence of this is that the early encounter stages of dwarf interactions would not result in a significant build-up of mass in the nuclei and instead result in a more distributed mass build-up.
This, combined with the dissipational nature of the gas and the high gas fraction of the progenitor galaxies suggests the remnant galaxy may retain or reform a dominant disk \citep[e.g.,][]{Barnes2004,Springel2005a,Robertson2006,Robertson2008} rather than becoming spheroid-dominated like mergers of massive galaxies.
Consistent with this scenario, the Illustris-TNG hydrodynamic cosmological simulations do not find that the morphology of galaxies with M$_{*}<10^{11}$ \Msun does not depend strongly on their merger history \citep{Rodriguez-Gomez2017}.

Simulations of low-mass galaxy mergers also show differences in the influence of mergers on low-mass galaxies, compared to more massive galaxies.
\citet{Kim2009} found distributed starbursts driven by shocks in high-resolution hydrodynamic simulations of low-mass, gas-rich galaxy mergers.
Including more detailed feedback treatment results in more well-defined starbursts in low-mass galaxies because the ISM becomes more strongly pressurized \citep{Hopkins2013}.
This is in contrast to massive galaxies where more realistic feedback treatment broadens and dilutes the starburst peak.
\citet{Besla2012} argue that starbursts should be preferentially introduced in the lower mass galaxy during dwarf-dwarf interactions.
This is consistent with the brightest \Ha knots and 60\% of the total \Ha flux residing to the lower-mass component of \source as well as its significantly higher sSFR.
Using cosmological zoom simulations \citet{Brooks2016} find that simulated low-mass merger remnants have less mass in their centers than the progenitor galaxies; while it is not possible to discern if this type of mass rearranging is occurring in \source, our observations are in qualitative agreement with their simulations that the remnants of low-mass mergers may not be as centrally concentrated as remnants of massive galaxy mergers.

In idealized merger simulations representing \emph{massive} high-redshift galaxies \citet{Fensch2017} found that higher gas fractions correspond to a reduced direct impact of the dynamical interaction on the star formation rate.
For both non-interacting and interacting galaxies, simulated star formation rates were elevated for systems with higher gas fractions.
However, the interacting system had fewer excursions above the main sequence of star formation and they were of reduced amplitude, compared to simulated interacting galaxies with lower gas fractions.
They argue strong inflows driven by clumps in the (non-interacting) high gas fraction disk result in a reduced effectiveness of tidally-induced torques to trigger the starburst, because gas inflows are already prevalent in those massive galaxies.

In comparing to low-redshift dwarf galaxies \citet{Fensch2017} note that, despite similar gas fractions to their simulations, non-interacting low-redshift dwarf galaxies have a reduced efficiency of inflows driven by disk instabilities and a more diffuse ISM at low-redshift.
They argue this makes dwarf galaxy interactions at low-redshift more likely to result in SFR enhancements (compared to their high-redshift simulations) by driving inflows far in excess of those secularly driven in disks.
However, we argue here that the enhancement in \source may not be the result of inflows, but rather large-scale ISM compression.
Determining whether torques are able to more efficiently drive gas to the center in later merger stages requires observations of other low-redshift, low-mass interacting systems.

\section{Conclusions}

We present VLT/MUSE IFU observations of the interacting dwarf pair \source.
We detect the major optical diagnostic lines, which we use to demonstrate the excitation of the ISM can be fully explained by star formation.
We also find that:
\begin{enumerate}
  \item{By combining the \Ha and $22~\mu m$ detections, we measure a total star formation rate of $0.48$ \Msun \pyr ($90\%$ of which is unobscured). This unobscured star formation is 2.7 times higher than inferred from extrapolations of SDSS spectroscopy.}
  \item{The specific star formation for the system ($5.3\times10^{-9}$ yr$^{-1}$) is more than an order of magnitude higher than the sSFR for non-interacting galaxies in the same mass range and yet is distributed across the source.
  The sSFR for the lower-mass galaxy of the pair shows an even larger enhancement, consistent with the theoretical expectations that starbursts should preferentially be trigger in the lower-mass member of dwarf pairs.}
  \item{The velocity structure, based on the \Ha emission, is not consistent with an ordered, rotating disk.}
  \item{Based on the kinematic and morphological evidence, and a ``lookalike'' numerical simulation, we conclude \source is a pair of dynamically interacting dwarf galaxies. This dynamical interaction is the likely cause of the substantially enhanced star formation.}
\end{enumerate}
This system suggests important hydrodynamical differences from more massive galaxy interactions, including more widespread star formation and a lack of large-scale shocks.
These SF properties may be explained by a scenario in which the pre-existing \HI disks/envelopes of dwarf galaxies are compressed in the interaction, which leads to overpressure and collapse of previously stable dense clumps of gas throughout the galaxy.
This suggests mergers affect the mass build-up of dwarf galaxies in a way that is different from massive galaxies.
In the future we will expand this study to other dwarf pairs in the TNT survey to assess how representative this object is of dwarf-dwarf interactions as class.

\acknowledgements

The authors thank the anonymous referee for their comments, which have improved the quality of the paper.

GCP was supported by a FONDECYT Postdoctoral Fellowship (No.\ 3150361).
DRP was supported by a Discovery Grant from NSERC of Canada.
NK is supported by the NSF CAREER award 1455260.
SEL acknowledges support from the NSF GRFP (Grant No.\ DDGE-1315231), the Virginia Space Grant Consortium, and the Clare Boothe Luce Foundation.

The authors thank F. E. Bauer for access to computing facilities which were used to process the MUSE data.
This research has made use of the NASA/IPAC Extragalactic Database (NED) and NASA's Astrophysics Data System.

Based on observations collected at the European Organisation for Astronomical Research in the Southern Hemisphere under ESO programme 097.B-0504(A).
This research has used data from the Sloan Digital Sky Survey.
Funding for the Sloan Digital Sky Survey IV has been provided by the Alfred P. Sloan Foundation, the U.S. Department of Energy Office of Science, and the Participating Institutions. SDSS acknowledges support and resources from the Center for High-Performance Computing at the University of Utah. The SDSS web site is www.sdss.org.
This publication makes use of data products from the Wide-field Infrared Survey Explorer, which is a joint project of the University of California, Los Angeles, and the Jet Propulsion Laboratory/California Institute of Technology, funded by the National Aeronautics and Space Administration.

\facility{VLT:Yepun (MUSE)}

\software{ESO Reflex \citep{Freudling2013}, ipython \citep{Perez2007}, numpy \citep{Vanderwalt2011}, matplotlib \citep{Hunter2007}, Astropy \citep{astropy}, Identikit \citep{Barnes2009,Barnes2010}, Zeno \citep{Barnes2011b}}

\bibliography{ms}

\begin{thebibliography}{}
\expandafter\ifx\csname natexlab\endcsname\relax\def\natexlab#1{#1}\fi
\providecommand{\url}[1]{\href{#1}{#1}}

\bibitem[{{Armus} {et~al.}(1989){Armus}, {Heckman}, \& {Miley}}]{Armus1989}
{Armus}, L., {Heckman}, T.~M., \& {Miley}, G.~K. 1989, \apj, 347, 727

\bibitem[{{Astropy Collaboration} {et~al.}(2013){Astropy Collaboration},
  {Robitaille}, {Tollerud}, {Greenfield}, {Droettboom}, {Bray}, {Aldcroft},
  {Davis}, {Ginsburg}, {Price-Whelan}, {Kerzendorf}, {Conley}, {Crighton},
  {Barbary}, {Muna}, {Ferguson}, {Grollier}, {Parikh}, {Nair}, {Unther},
  {Deil}, {Woillez}, {Conseil}, {Kramer}, {Turner}, {Singer}, {Fox}, {Weaver},
  {Zabalza}, {Edwards}, {Azalee Bostroem}, {Burke}, {Casey}, {Crawford},
  {Dencheva}, {Ely}, {Jenness}, {Labrie}, {Lim}, {Pierfederici}, {Pontzen},
  {Ptak}, {Refsdal}, {Servillat}, \& {Streicher}}]{astropy}
{Astropy Collaboration}, {Robitaille}, T.~P., {Tollerud}, E.~J., {et~al.} 2013,
  \aap, 558, A33

\bibitem[{{Bacon} {et~al.}(2010){Bacon}, {Accardo}, {Adjali}, {Anwand},
  {Bauer}, {Biswas}, {Blaizot}, {Boudon}, {Brau-Nogue}, {Brinchmann},
  {Caillier}, {Capoani}, {Carollo}, {Contini}, {Couderc}, {Daguis{\'e}},
  {Deiries}, {Delabre}, {Dreizler}, {Dubois}, {Dupieux}, {Dupuy}, {Emsellem},
  {Fechner}, {Fleischmann}, {Fran{\c c}ois}, {Gallou}, {Gharsa}, {Glindemann},
  {Gojak}, {Guiderdoni}, {Hansali}, {Hahn}, {Jarno}, {Kelz}, {Koehler},
  {Kosmalski}, {Laurent}, {Le Floch}, {Lilly}, {Lizon}, {Loupias}, {Manescau},
  {Monstein}, {Nicklas}, {Olaya}, {Pares}, {Pasquini}, {P{\'e}contal-Rousset},
  {Pell{\'o}}, {Petit}, {Popow}, {Reiss}, {Remillieux}, {Renault}, {Roth},
  {Rupprecht}, {Serre}, {Schaye}, {Soucail}, {Steinmetz}, {Streicher}, {Stuik},
  {Valentin}, {Vernet}, {Weilbacher}, {Wisotzki}, \& {Yerle}}]{Bacon2010}
{Bacon}, R., {Accardo}, M., {Adjali}, L., {et~al.} 2010, in \procspie, Vol.
  7735, Ground-based and Airborne Instrumentation for Astronomy III, 773508

\bibitem[{{Baldwin} {et~al.}(1981){Baldwin}, {Phillips}, \&
  {Terlevich}}]{Baldwin1981}
{Baldwin}, J.~A., {Phillips}, M.~M., \& {Terlevich}, R. 1981, \pasp, 93, 5

\bibitem[{Barnes(1988)}]{Barnes1988}
Barnes, J.~E. 1988, \apj, 331, 699

\bibitem[{Barnes(2004)}]{Barnes2004}
---. 2004, \mnras, 350, 798

\bibitem[{{Barnes}(2011)}]{Barnes2011b}
{Barnes}, J.~E. 2011, ZENO: N-body and SPH Simulation Codes, Astrophysics
  Source Code Library, ,

\bibitem[{Barnes \& Hernquist(1996)}]{Barnes1996}
Barnes, J.~E., \& Hernquist, L. 1996, \apj, 471, 115

\bibitem[{Barnes \& Hernquist(1991)}]{Barnes1991}
Barnes, J.~E., \& Hernquist, L.~E. 1991, \apj, 370, L65

\bibitem[{Barnes \& Hibbard(2009)}]{Barnes2009}
Barnes, J.~E., \& Hibbard, J.~E. 2009, \aj, 137, 3071

\bibitem[{{Barnes} \& {Hibbard}(2010)}]{Barnes2010}
{Barnes}, J.~E., \& {Hibbard}, J.~E. 2010, Identikit 1: A Modeling Tool for
  Interacting Disk Galaxies, Astrophysics Source Code Library, , ,
  ascl:1011.001

\bibitem[{Besla {et~al.}(2012)Besla, Kallivayalil, Hernquist, van~der Marel,
  Cox, \& Kere\v{s}}]{Besla2012}
Besla, G., Kallivayalil, N., Hernquist, L., {et~al.} 2012, \mnras, 421, 2109

\bibitem[{{Bradford} {et~al.}(2015){Bradford}, {Geha}, \&
  {Blanton}}]{Bradford2015}
{Bradford}, J.~D., {Geha}, M.~C., \& {Blanton}, M.~R. 2015, \apj, 809, 146

\bibitem[{{Brinchmann} {et~al.}(2004){Brinchmann}, {Charlot}, {White},
  {Tremonti}, {Kauffmann}, {Heckman}, \& {Brinkmann}}]{Brinchmann2004}
{Brinchmann}, J., {Charlot}, S., {White}, S.~D.~M., {et~al.} 2004, \mnras, 351,
  1151

\bibitem[{{Brooks} \& {Christensen}(2016)}]{Brooks2016}
{Brooks}, A., \& {Christensen}, C. 2016, Bulge Formation via Mergers in
  Cosmological Simulations, Vol. 418, 317

\bibitem[{{Cappellari} \& {Copin}(2003)}]{Cappellari2003}
{Cappellari}, M., \& {Copin}, Y. 2003, \mnras, 342, 345

\bibitem[{{Chang} {et~al.}(2015){Chang}, {van der Wel}, {da Cunha}, \&
  {Rix}}]{Chang2015a}
{Chang}, Y.-Y., {van der Wel}, A., {da Cunha}, E., \& {Rix}, H.-W. 2015, \apjs,
  219, 8

\bibitem[{{Duc} {et~al.}(2015){Duc}, {Cuillandre}, {Karabal}, {Cappellari},
  {Alatalo}, {Blitz}, {Bournaud}, {Bureau}, {Crocker}, {Davies}, {Davis}, {de
  Zeeuw}, {Emsellem}, {Khochfar}, {Krajnovic}, {Kuntschner}, {McDermid},
  {Michel-Dansac}, {Morganti}, {Naab}, {Oosterloo}, {Paudel}, {Sarzi}, {Scott},
  {Serra}, {Weijmans}, \& {Young}}]{Duc2015}
{Duc}, P.-A., {Cuillandre}, J.-C., {Karabal}, E., {et~al.} 2015, \mnras, 446,
  120

\bibitem[{{Fensch} {et~al.}(2017){Fensch}, {Renaud}, {Bournaud}, {Duc},
  {Agertz}, {Amram}, {Combes}, {Di Matteo}, {Elmegreen}, {Emsellem}, {Jog},
  {Perret}, {Struck}, \& {Teyssier}}]{Fensch2017}
{Fensch}, J., {Renaud}, F., {Bournaud}, F., {et~al.} 2017, \mnras, 465, 1934

\bibitem[{{Freudling} {et~al.}(2013){Freudling}, {Romaniello}, {Bramich},
  {Ballester}, {Forchi}, {Garc{\'{\i}}a-Dabl{\'o}}, {Moehler}, \&
  {Neeser}}]{Freudling2013}
{Freudling}, W., {Romaniello}, M., {Bramich}, D.~M., {et~al.} 2013, \aap, 559,
  A96

\bibitem[{{Gordon} {et~al.}(2003){Gordon}, {Clayton}, {Misselt}, {Landolt}, \&
  {Wolff}}]{Gordon2003}
{Gordon}, K.~D., {Clayton}, G.~C., {Misselt}, K.~A., {Landolt}, A.~U., \&
  {Wolff}, M.~J. 2003, \apj, 594, 279

\bibitem[{{Hinshaw} {et~al.}(2013){Hinshaw}, {Larson}, {Komatsu}, {Spergel},
  {Bennett}, {Dunkley}, {Nolta}, {Halpern}, {Hill}, {Odegard}, {Page}, {Smith},
  {Weiland}, {Gold}, {Jarosik}, {Kogut}, {Limon}, {Meyer}, {Tucker}, {Wollack},
  \& {Wright}}]{Hinshaw2013}
{Hinshaw}, G., {Larson}, D., {Komatsu}, E., {et~al.} 2013, \apjs, 208, 19

\bibitem[{{Hopkins} {et~al.}(2013){Hopkins}, {Cox}, {Hernquist}, {Narayanan},
  {Hayward}, \& {Murray}}]{Hopkins2013}
{Hopkins}, P.~F., {Cox}, T.~J., {Hernquist}, L., {et~al.} 2013, \mnras, 430,
  1901

\bibitem[{{Hopkins} {et~al.}(2008){Hopkins}, {Hernquist}, {Cox}, \& {Kere{\v
  s}}}]{Hopkins2008}
{Hopkins}, P.~F., {Hernquist}, L., {Cox}, T.~J., \& {Kere{\v s}}, D. 2008,
  \apjs, 175, 356

\bibitem[{Howell {et~al.}(2010)Howell, Armus, Mazzarella, Evans, Surace,
  Sanders, Petric, Appleton, Bothun, Bridge, Chan, Charmandaris, Frayer, Haan,
  Inami, Kim, Lord, Madore, Melbourne, Schulz, U, Vavilkin, Veilleux, \&
  Xu}]{Howell2010}
Howell, J.~H., Armus, L., Mazzarella, J.~M., {et~al.} 2010, \apj, 715, 572

\bibitem[{Hunter(2007)}]{Hunter2007}
Hunter, J.~D. 2007, Computing In Science \& Engineering, 9, 90

\bibitem[{Jog \& Solomon(1992)}]{Jog1992}
Jog, C.~J., \& Solomon, P.~M. 1992, \apj, 387, 152

\bibitem[{Kennicutt(1998)}]{Kennicutt1998}
Kennicutt, R.~C. 1998, \araa, 36, 189

\bibitem[{{Kewley} {et~al.}(2006){Kewley}, {Groves}, {Kauffmann}, \&
  {Heckman}}]{Kewley2006}
{Kewley}, L.~J., {Groves}, B., {Kauffmann}, G., \& {Heckman}, T. 2006, \mnras,
  372, 961

\bibitem[{{Kim} {et~al.}(2009){Kim}, {Wise}, \& {Abel}}]{Kim2009}
{Kim}, J.-h., {Wise}, J.~H., \& {Abel}, T. 2009, \apjl, 694, L123

\bibitem[{{Lee} {et~al.}(2009{\natexlab{a}}){Lee}, {Kennicutt}, {Funes},
  {Sakai}, \& {Akiyama}}]{Lee2009a}
{Lee}, J.~C., {Kennicutt}, Jr., R.~C., {Funes}, S.~J.~J.~G., {Sakai}, S., \&
  {Akiyama}, S. 2009{\natexlab{a}}, \apj, 692, 1305

\bibitem[{{Lee} {et~al.}(2009{\natexlab{b}}){Lee}, {Gil de Paz}, {Tremonti},
  {Kennicutt}, {Salim}, {Bothwell}, {Calzetti}, {Dalcanton}, {Dale},
  {Engelbracht}, {Funes}, {Johnson}, {Sakai}, {Skillman}, {van Zee}, {Walter},
  \& {Weisz}}]{Lee2009}
{Lee}, J.~C., {Gil de Paz}, A., {Tremonti}, C., {et~al.} 2009{\natexlab{b}},
  \apj, 706, 599

\bibitem[{{Lee} {et~al.}(2011){Lee}, {Gil de Paz}, {Kennicutt}, {Bothwell},
  {Dalcanton}, {Jos{\'e} G.~Funes S.}, {Johnson}, {Sakai}, {Skillman},
  {Tremonti}, \& {van Zee}}]{Lee2011}
{Lee}, J.~C., {Gil de Paz}, A., {Kennicutt}, Jr., R.~C., {et~al.} 2011, \apjs,
  192, 6

\bibitem[{{Maraston} {et~al.}(2009){Maraston}, {Str{\"o}mb{\"a}ck}, {Thomas},
  {Wake}, \& {Nichol}}]{Maraston2006}
{Maraston}, C., {Str{\"o}mb{\"a}ck}, G., {Thomas}, D., {Wake}, D.~A., \&
  {Nichol}, R.~C. 2009, \mnras, 394, L107

\bibitem[{Mihos \& Hernquist(1994)}]{Mihos1994a}
Mihos, J.~C., \& Hernquist, L. 1994, \apj, 431, L9

\bibitem[{{Moreno} {et~al.}(2015){Moreno}, {Torrey}, {Ellison}, {Patton},
  {Bluck}, {Bansal}, \& {Hernquist}}]{Moreno2015}
{Moreno}, J., {Torrey}, P., {Ellison}, S.~L., {et~al.} 2015, \mnras, 448, 1107

\bibitem[{{Osterbrock} \& {Ferland}(2006)}]{Osterbrock2006}
{Osterbrock}, D.~E., \& {Ferland}, G.~J. 2006, Astrophysics of gaseous nebulae
  and active galactic nuclei

\bibitem[{{Patton} {et~al.}(2013){Patton}, {Torrey}, {Ellison}, {Mendel}, \&
  {Scudder}}]{Patton2013}
{Patton}, D.~R., {Torrey}, P., {Ellison}, S.~L., {Mendel}, J.~T., \& {Scudder},
  J.~M. 2013, \mnras, 433, L59

\bibitem[{{Pearson} {et~al.}(2016){Pearson}, {Besla}, {Putman}, {Lutz},
  {Fernandez}, {Stierwalt}, {Patton}, {Kim}, {Kallivayalil}, {Johnson}, \&
  {Sung}}]{Pearson2016}
{Pearson}, S., {Besla}, G., {Putman}, M.~E., {et~al.} 2016, \mnras, 459, 1827

\bibitem[{P\'erez \& Granger(2007)}]{Perez2007}
P\'erez, F., \& Granger, B.~E. 2007, Computing in Science and Engineering, 9,
  21

\bibitem[{Privon {et~al.}(2013)Privon, Barnes, Evans, Hibbard, Yun, Mazzarella,
  Armus, \& Surace}]{Privon2013}
Privon, G.~C., Barnes, J.~E., Evans, A.~S., {et~al.} 2013, \apj, 771, 120

\bibitem[{{Robertson} {et~al.}(2006){Robertson}, {Bullock}, {Cox}, {Di Matteo},
  {Hernquist}, {Springel}, \& {Yoshida}}]{Robertson2006}
{Robertson}, B., {Bullock}, J.~S., {Cox}, T.~J., {et~al.} 2006, \apj, 645, 986

\bibitem[{Robertson \& Bullock(2008)}]{Robertson2008}
Robertson, B.~E., \& Bullock, J.~S. 2008, \apj, 685, L27

\bibitem[{{Rodriguez-Gomez} {et~al.}(2017){Rodriguez-Gomez}, {Sales}, {Genel},
  {Pillepich}, {Zjupa}, {Nelson}, {Griffen}, {Torrey}, {Snyder},
  {Vogelsberger}, {Springel}, {Ma}, \& {Hernquist}}]{Rodriguez-Gomez2017}
{Rodriguez-Gomez}, V., {Sales}, L.~V., {Genel}, S., {et~al.} 2017, \mnras, 467,
  3083

\bibitem[{Sanders {et~al.}(1988)Sanders, Soifer, Elias, Madore, Matthews,
  Neugebauer, \& Scoville}]{Sanders1988}
Sanders, D.~B., Soifer, B.~T., Elias, J.~H., {et~al.} 1988, \apj, 325, 74

\bibitem[{Springel \& Hernquist(2005)}]{Springel2005a}
Springel, V., \& Hernquist, L. 2005, \apj, 622, L9

\bibitem[{{Stierwalt} {et~al.}(2015){Stierwalt}, {Besla}, {Patton}, {Johnson},
  {Kallivayalil}, {Putman}, {Privon}, \& {Ross}}]{Stierwalt2015}
{Stierwalt}, S., {Besla}, G., {Patton}, D., {et~al.} 2015, \apj, 805, 2

\bibitem[{{Stierwalt} {et~al.}(2017){Stierwalt}, {Liss}, {Johnson}, {Patton},
  {Privon}, {Besla}, {Kallivayalil}, \& {Putman}}]{Stierwalt2017}
{Stierwalt}, S., {Liss}, S.~E., {Johnson}, K.~E., {et~al.} 2017, Nature
  Astronomy, 1, 0025

\bibitem[{Stierwalt {et~al.}(2013)Stierwalt, Armus, Surace, Inami, Petric,
  D\'{\i}az-Santos, Haan, Charmandaris, Howell, Kim, Marshall, Mazzarella,
  Spoon, Veilleux, Evans, Sanders, Appleton, Bothun, Bridge, Chan, Frayer,
  Iwasawa, Kewley, Lord, Madore, Melbourne, Murphy, Rich, Schulz, Sturm, U,
  Vavilkin, \& Xu}]{Stierwalt2013}
Stierwalt, S., Armus, L., Surace, J.~A., {et~al.} 2013, \apj, 206, 1

\bibitem[{{Swaters} {et~al.}(2002){Swaters}, {van Albada}, {van der Hulst}, \&
  {Sancisi}}]{Swaters2002}
{Swaters}, R.~A., {van Albada}, T.~S., {van der Hulst}, J.~M., \& {Sancisi}, R.
  2002, \aap, 390, 829

\bibitem[{Toomre \& Toomre(1972)}]{Toomre1972}
Toomre, A., \& Toomre, J. 1972, \apj, 178, 623

\bibitem[{{Tremonti} {et~al.}(2004){Tremonti}, {Heckman}, {Kauffmann},
  {Brinchmann}, {Charlot}, {White}, {Seibert}, {Peng}, {Schlegel}, {Uomoto},
  {Fukugita}, \& {Brinkmann}}]{Tremonti2004}
{Tremonti}, C.~A., {Heckman}, T.~M., {Kauffmann}, G., {et~al.} 2004, \apj, 613,
  898

\bibitem[{{Van Der Walt} {et~al.}(2011){Van Der Walt}, {Colbert}, \&
  {Varoquaux}}]{Vanderwalt2011}
{Van Der Walt}, S., {Colbert}, S.~C., \& {Varoquaux}, G. 2011, ArXiv e-prints,
  arXiv:1102.1523

\bibitem[{{Veilleux} \& {Osterbrock}(1987)}]{Veilleux1987}
{Veilleux}, S., \& {Osterbrock}, D.~E. 1987, \apjs, 63, 295

\bibitem[{{Wright} {et~al.}(2010){Wright}, {Eisenhardt}, {Mainzer}, {Ressler},
  {Cutri}, {Jarrett}, {Kirkpatrick}, {Padgett}, {McMillan}, {Skrutskie},
  {Stanford}, {Cohen}, {Walker}, {Mather}, {Leisawitz}, {Gautier}, {McLean},
  {Benford}, {Lonsdale}, {Blain}, {Mendez}, {Irace}, {Duval}, {Liu}, {Royer},
  {Heinrichsen}, {Howard}, {Shannon}, {Kendall}, {Walsh}, {Larsen}, {Cardon},
  {Schick}, {Schwalm}, {Abid}, {Fabinsky}, {Naes}, \& {Tsai}}]{Wright2010}
{Wright}, E.~L., {Eisenhardt}, P.~R.~M., {Mainzer}, A.~K., {et~al.} 2010, \aj,
  140, 1868

\end{thebibliography}

\end{document}